\documentclass[showpacs,aps,pre,twocolumn,superscriptaddress]{revtex4} 
\usepackage{graphicx}     
\usepackage{bm}
\begin{document}
\title{Finite difference lattice Boltzmann model with flux limiters for
liquid-vapor systems} 
\author{V. Sofonea}
\email[]{sofonea@acad-tim.tm.edu.ro}
\affiliation{Dipartimento di Fisica, Universit\`{a} di Bari,
Via Amendola 173, 70126 Bari, Italy}
\affiliation{Laboratory for Numerical Simulation and Parallel Computing
in Fluid Mechanics,\\Center for Fundamental and Advanced Technical Research,
Romanian Academy\\Bd. Mihai Viteazul 24, 300223 Timi\c soara, 
Romania}
\altaffiliation{Permanent address}
\author{A. Lamura}
\email[]{a.lamura@area.ba.cnr.it}
\homepage[]{http://www.ba.cnr.it/~irmaal21}
\affiliation{Istituto Applicazioni Calcolo, CNR, Sezione di Bari,
Via Amendola 122/D, 70126 Bari, Italy}
\author{G. Gonnella}
\email[]{gonnella@ba.infn.it}
\affiliation{Dipartimento di Fisica, Universit\`{a} di Bari,
Via Amendola 173, 70126 Bari, Italy}
\affiliation{TIRES, Center of Innovative Technologies for Signal Detection and
Processing,
{\it and} INFM,
Unit\`{a} di Bari,
{\it and} INFN, Sezione di Bari,
Via Amendola 173, 70126 Bari, Italy}
\author{A. Cristea}
\email[]{f1astra@acad-tim.tm.edu.ro}
\affiliation{Laboratory for Numerical Simulation and Parallel Computing
in Fluid Mechanics,\\Center for Fundamental and Advanced Technical Research,
Romanian Academy\\Bd. Mihai Viteazul 24, 300223 Timi\c soara, 
Romania}
\date{\today}
\begin{abstract}
In this paper we apply a finite difference lattice 
Boltzmann model to study the phase separation in a two-dimensional
liquid-vapor system.
Spurious numerical effects in macroscopic equations
are discussed  and
an appropriate numerical scheme involving flux limiter techniques
is proposed to minimize them and guarantee a better numerical stability
at very low viscosity.
The phase separation kinetics is investigated and we find evidence of two
different growth regimes depending on the value of the fluid viscosity
as well as on the liquid-vapor ratio.
\end{abstract}
\maketitle

\section{Introduction}

Lattice Boltzmann
(LB) models approach physical phenomena in
fluid systems using a phase-space discretized form of the Boltzmann equation
\cite{benzi1992,rotzal,chopardroz,gladrow,succibook}. Conservation equations
are derived by calculating moments of various order of this equation
\cite{grad,chapmancowling,vincenti,burgers,huang,gombosi,sosek2001}.
After the publication of the first LB model which exhibits phase separation
\cite{shanchen1993,shanchen1994},
LB models were widely used to investigate the complex behavior
of single- or multi-component/phase fluid systems \cite{chopardroz,succibook}
and refer mainly to isothermal systems
\cite{succi,gunstensen1991,swift1995,orlandini1995,
swift1996,doolen,heshandoolen1998,lamura,rector,kalarakis,theofanous2003}. This
limitation
comes from the
constant value of the lattice speed $c_l$ which in LB models
is related to the temperature $T$, the lattice
spacing $\delta s$ and the time step $\delta t$ through two separate
relations 
\begin{eqnarray}
c_l & = & \frac{c_s}{\sqrt{\chi}} = 
\sqrt{\,\frac{\,k_B T\,}{\,\chi m}\,} \label{cthermal}\\
c_l & = & \frac{\,\delta s\,}{\,\delta t\,} \rule{0mm}{8mm} \label{clattice}
\end{eqnarray}
where $c_s = \sqrt{k_B T / m}$ is the isothermal 
speed of sound for an ideal fluid, 
$m$ is the mass of fluid particles, 
$\chi$ is a constant depending on the geometry of the lattice,
and $k_B$ is Boltzmann's constant \cite{succibook,heluo}.

According to the ``collide and stream'' philosophy of
LB models, fluid collides in the lattice nodes and
thereafter moves along the lattice links
in a lapse $\delta t$ towards neighboring nodes with the
speed $c_l$ given by  Eq.~(\ref{clattice})
\cite{rotzal,chopardroz,gladrow,succibook}.
Such a relationship is no longer considered in
finite difference lattice Boltzmann (FDLB) models
\cite{cao,weishyy,seta,leelin,sosek2003} which start directly from the
Boltzmann equation and have a better numerical stability. 
In such models there is 
more freedom to choose the discrete velocity set,
as done recently in the thermal FDLB model of Watari and 
Tsutahara \cite{watari}
where the possibility of having different sets of velocities allows
to release the constraint of constant temperature.
Also, the use of FDLB models
is promising, e.g., when
considering LB models for multicomponent fluid systems, where the
masses of the components are not identical and
Eq.~(\ref{cthermal}) would lead to different lattice speeds. 
In this context, FDLB 
models may be viewed as a convenient alternative to interpolation supplemented
LB models \cite{he1996,he1997,hipc2003}. 

FDLB models, as well as LB models,
are known to introduce spurious terms in the mass and
momentum conservation equations, which are dependent on the lattice
spacing $\delta s$ and the time step $\delta t$ \cite{sosek2003}.
The behavior of an isothermal fluid system subjected to FDLB simulation
is governed by the apparent values of the viscosity and/or diffusivity. The
expression of these quantities with respect to $\delta s$ and $\delta t$
depends on the finite difference scheme used in the FDLB model. Consequently,
the choice of the numerical scheme may alter significantly the macroscopic
behavior of the fluid system observed during simulations as well
as the numerical stability. This problem
still lacks necessary clarification and should be always considered
in order to recover the correct physical interpretation of simulation
results.

The purpose of this paper is to investigate these numerical aspects
by using a FDLB model
addressing the phase separation kinetics in a van der Waals fluid. 
Phase separation in liquid-vapor systems has not received as much attention
as in binary fluids \cite{yeom}. 
Under the hypothesis of dynamical scaling 
the late time kinetics can be characterized in terms of a single 
length scale $R(t)$ which grows according to the power law
$R(t) \sim t^{\alpha}$, where $\alpha$ is the growth exponent \cite{bray}.
The late time growth, when hydrodynamics is neglected, is expected
to be described by the Allen-Cahn theory which gives a growth exponent 
$\alpha = 1/2$ \cite{allen}. When hydrodynamics comes into play, the 
liquid-vapor system behaves similarly to binary fluids so that
a growth exponent $\alpha= 2/3$ is expected \cite{yeom}.
Previous numerical studies used molecular dynamics simulations 
\cite{yama1,yama2}
and a LB model based on a free energy functional 
\cite{osborn,mecke1,mecke2}.
In molecular dynamics simulations 
it was found evidence for the 
growth exponent $1/2$ \cite{yama1,yama2}.
By using 
the free-energy LB model, Osborn {\it et al.} \cite{osborn} 
found the growth exponents
$2/3$ and $1/2$ at low and high viscosity, respectively, independently
on the system composition. 
Mecke and Sofonea \cite{mecke1,mecke2}, using the same algorithm for
an off-symmetric system,
found a crossover from $2/3$
to $1/2$ at low viscosity, and $1/3$ at high viscosity. 
We will compare results of our model with the aforementioned ones.

To model the liquid-vapor system, a standard
force term \cite{shanchen1993,shanchen1994,theofanous2003} 
is added to the discretized Boltzmann
equations. The resulting FDLB model is described in Section
\ref{secdescription}. 
In Section \ref{secfdschemes} we introduce two numerical
schemes, namely, the first order upwind finite difference scheme and
a higher order one which uses flux limiters \cite{leveque,toro}.
There we show the difference between FDLB, ``collide and stream'' LB and
volumetric LB models \cite{chen1998,chen2001}.
We thereafter discuss the spurious numerical effects these schemes introduce in
the fluid equations. Section \ref{secresults} reports the
simulation results, where special attention was given to 
the effects of the numerical schemes on estimation of the growth exponent.
In order to clarify the phenomenology and estimate accurately
the exponent $\alpha$, we monitored the size of domains $R(t)$ by using three
independent measures. 
A discussion about the method and results ends the paper.

\section{\label{secdescription}The model}

The $2D$ FDLB model follows the LB model for
non-ideal fluids 
\cite{shanchen1993,shanchen1994,heshandoolen1998,luoprl98,luopre00}.
The starting point is provided by the set of\, ${\mathcal{N}}$ partial
derivatives equations resulting from the discretization of the Boltzmann
equation on a square lattice ${\mathcal{L}}$ when the collision term is
linearized using the BGK approximation \cite{bgk}.
In non-dimensional form, this set reads
\begin{eqnarray}
\partial_{t}f_i+e_{i\beta}\partial_{\beta}f_i & = &
\frac{1}{\chi c^2}f_{i}^{eq}(e_{i\beta}-u_{\beta})F_{\beta}
-\frac{1}{\tau}(f_{i}-f_{i}^{eq})
\nonumber\\
& & i = 0,1,\ldots {\mathcal{N}}\label{lbeqs}
\end{eqnarray}
Since we will deal with a van der Waals fluid,
we used the following reference quantities for particle number
density, temperature and speed to get the non-dimensional form (\ref{lbeqs})
of the discretized Boltzmann equations: $n_{R}=N_{A}/V_{mc}$, $T_{R}=T_{c}$,
$c_{R}=\sqrt{k_{B}T_{c}/m}$. Here $N_{A}$ is Avogadro's number,
$V_{mc}$ is the molar volume at the critical point and $T_{c}$ is the
critical temperature. With this choice of reference quantities, the
dimensionless speed is \cite{ijmpc03}
\begin{equation}
c=c_l/c_R=\sqrt{\theta/\chi}
\label{cdef}
\end{equation}
where $\theta=T/T_R$ is the
dimensionless temperature and the constant $\chi$ equals $1/3$
for the square lattice we use (see later for details on the lattice)
\cite{sosek2003}. If we take the system size as 
the reference length $l_{R}$, we get the reference time $t_{R}$ from the
condition $t_{R}c_{R}=l_{R}$. The non-dimensionalized lattice spacing
is defined by the number of lattice nodes $N$ we
choose along the non-dimensionalized system length $L$
\begin{equation}
\delta s = \frac{L}{N}
\label{deltasdef}
\end{equation}

The particle distribution functions $f_{i}\equiv f_{i}({\bm{x}},t)$
are defined in the nodes ${\bm{x}}$ of the square lattice ${\mathcal{L}}$.
In the $D2Q9$ model we use in this paper,
${\mathcal{N}}=8$ and the 
velocities ${\bm{e}}_{i}$ are \cite{rotzal,chopardroz,gladrow,succibook}
\begin{eqnarray}
{\bm{e}}_{0} & = & 0 \nonumber\\
{\bm{e}}_{i} & = & \left[\,\cos\frac{\pi(i-1)}{2}\, ,\,\sin\frac{\pi(i-1)}{2}\,
\right] c \rule{0mm}{8mm}
\label{eidef} \\ & & (i\,=\,1,\dots 4) \rule{0mm}{5mm}\nonumber \\
{\bm{e}}_{i} & = & 
\left[\,\cos\left(\frac{\pi}{4}\,+\,\frac{\pi(i-5)}{2}\right)\,
,\,\sin\left(\frac{\pi}{4}\,+\,\frac{\pi(i-5)}{2}\right)
\right] c\sqrt{2} \rule{0mm}{8mm}\nonumber \\* & & (i\,=\,5,\dots 8)
\rule{0mm}{5mm} \nonumber
\end{eqnarray}

The equilibrium distribution functions $f_{i}^{eq}=f_i^{eq}({\bm{x}},t)$
are expressed as series expansions of the Maxwellian
distribution function, up to second order with respect to the
local velocity ${\bm{u}}={\bm{u}}({\bm{x}},t)$, whose Cartesian components are
$u_{\beta}$ \cite{qian}:
\begin{equation}
f_{i}^{eq}  = w_i\,n \left[ 1 + \frac{ {\bm{e}}_{i}\cdot
{\bm{u}}}{\chi c^2} + \frac{({\bm{e}}_{i}\cdot{\bm{u}})^2}{2{\chi}^2 c^4} -
\frac{({\bm{u}})^2}{2{\chi} c^2} \right]
\label{feqdef}
\end{equation}
The weight coefficients are:
\begin{equation}\label{weights}
w_i\,=\,\left\{\,\begin{array}{ll}
\frac{\,4\,}{\,9\,} & (i\,=\,0)\\
\frac{\,1\,}{\,9\,} & (i\,=\,1,\dots 4) \rule{0mm}{7mm}\\
\frac{\,1\,}{\,36\,} & (i\,=\,5,\dots 8)\rule{0mm}{7mm}
\end{array}\right.
\end{equation}
The local density $n=n({\bm{x}},t)$, as well as the components
of the local velocity ${\bm{u}}$ which enter Eq.~(\ref{feqdef}), are calculated
from the distribution functions as follows:
\begin{eqnarray}
n & = & \sum_{i} f_{i} = \sum_{i} f_{i}^{eq} \\
u_{\beta} & = & \frac{1}{n}\sum_{i} f_{i}e_{i\beta} =
\frac{1}{n} \sum_{i} f_{i}^{eq} e_{i\beta}
\end{eqnarray}

The force term in Eqs.~(\ref{lbeqs}) is given by
\cite{luoprl98,luopre00} 
\begin{equation}
F_{\beta} = \frac{1}{n}\partial_\beta (p^{i} - p^{w}) +
\kappa\partial_\beta (\nabla^2 n)
\label{forcedef}
\end{equation}
where 
\begin{equation}
p^{i} = \theta n
\end{equation}
and
\begin{equation}
p^{w} = \frac{3\theta n}{3-n}-\frac{9}{8}n^2 
\label{vdw}
\end{equation}
are the non-dimensionalized pressures of the
ideal and the van der Waals fluid, respectively \cite{ijmpc03}.
With the equation of state in the form (\ref{vdw}), 
the critical point is located at
$\theta=1$ and $n=1$.
The parameter $\kappa$ controls the surface tension \cite{theofanous2003}.
The mass and momentum equations are recovered from Eqs.~(\ref{lbeqs}) after
using the standard Chapman - Enskog procedure up to second order with respect
to Knudsen number $Kn = c \tau / L$. 
These equations read
\cite{gladrow,succibook,theofanous2003}
\begin{eqnarray}
\partial_t n + \partial_{\beta}(nu_{\beta}) & = & 0
\label{physmass}\\
\partial_t(nu_{\alpha}) + \partial_{\beta}(nu_{\alpha}u_{\beta}) & = &
-\partial_{\alpha}p^w + \kappa n\partial_{\alpha}(\nabla^2 n)
\rule{0mm}{5mm}\label{physmom} \\ 
& + & \nu\partial_{\beta}\left[n\left(\partial_{\alpha}u_{\beta}+
\partial_{\beta}u_{\alpha}\right)\right] \rule{0mm}{5mm} \nonumber
\end{eqnarray}
where
\begin{equation}
\nu = \chi c^2\tau
\label{physnu}
\end{equation}
is the {\emph{physical value}} of the kinematic viscosity \cite{sosek2003}.
The particular numerical scheme used to solve Eqs.~(\ref{lbeqs})
may introduce a spurious viscosity term that adds to the physical value,
as seen in the next section.
Finally, we note that 
the force term (\ref{forcedef}) allows to recover the Navier - Stokes
equation (\ref{physmom})
where the pressure $p^w$ appearing on the r.h.s. 
is subjected to the van der Waals equation of state (\ref{vdw}).

\section{\label{secfdschemes}Finite difference schemes}

\subsection{First-order upwind scheme}

The set of phase space discretized equations (\ref{lbeqs}) may be solved
numerically by using an appropriate finite difference scheme defined
on the lattice ${\mathcal{L}}$. Simple second-order schemes like the centered
one or the Lax - Wendroff scheme \cite{leveque,toro,leveque2002} are unstable
because of large values of the density gradient which may occur in the
interface regions separating the liquid and vapor phases of the van der Waals
fluid. The first-order upwind scheme, which is also used in LB
models \cite{rotzal,chopardroz,gladrow,succibook}, is a good candidate because
of its stability. When associated to the forward time stepping rule,
this scheme gives the following updating rule for the distribution functions
defined in node ${\bm{x}}\in{\mathcal{L}}$ \cite{sosek2003}
\begin{eqnarray}
& & f_i({\bm{x}},t+\delta t) \, = \, f_i({\bm{x}},t) - \label{uprule}\\
& & \frac{c\delta t}{\delta s}
\left[f_i({\bm{x}},t)-f_i({\bm{x}}-\delta s{\bm{e}}_i/c,t)\right] +
\delta t Q_{i}({\bm{x}},t) \rule{0mm}{7mm}\nonumber\\
\end{eqnarray}
where
\begin{eqnarray}
& & Q_{i} = Q_{i}({\bm{x}},t) = \frac{1}{\theta}\times \rule{0mm}{7mm}\\
& & \left\{\frac{1}{n({\bm{x}},t)}
\partial_{\beta}\left[p^i({\bm{x}},t)-p^w({\bm{x}},t)\right]+
\kappa\partial_{\beta}\left[\nabla^{2} n({\bm{x}},t)\right]\right\} 
\rule{0mm}{7mm}\nonumber\\
& & \times f_{i}^{eq}({\bm{x}},t)
\left[e_{i\beta}-u_{\beta}({\bm{x}},t)\right] - \frac{1}{\tau}
\left[f_{i}({\bm{x}},t)-f_{i}^{eq}({\bm{x}},t)\right] ,
\rule{0mm}{7mm}\nonumber\\
& & i = 0,1,\ldots {\mathcal{N}} \rule{0mm}{6mm}\nonumber
\end{eqnarray}

As discussed in Ref.~\cite{sosek2003}, 
finite difference schemes introduce spurious
numerical terms in the conservation equations. This happens because the
{\emph{real}} evolution equations recovered (up to second order in space and
time) from the updating rules (\ref{uprule}) are
\begin{eqnarray}
\partial_{t}f_i+\phi\partial_{t}^2 f_{i}+e_{i\beta}\partial_{\beta}f_i
-\psi\partial_{\beta}\partial_{\gamma}e_{i\beta}e_{i\gamma}f_{i} & = &
Q_{i} ,\nonumber\\
i = 0,1,\ldots {\mathcal{N}}  & & \rule{0mm}{5mm}
\label{lbeqsreal}
\end{eqnarray}
where
\begin{eqnarray}
\phi & = & \frac{\delta t}{2} \label{phidef}\\
\psi & = & \frac{\delta s}{2c} \rule{0mm}{6mm}\label{psidef}
\end{eqnarray}
We get the following form of the conservation equations up to second
order in the Knudsen number:
\begin{equation}
\partial_t n + \partial_{\beta}(nu_{\beta})  =  
(\psi - \phi) \partial_{\alpha} \partial_{\beta}
\left [ \chi c^2 n \delta_{\alpha \beta} + n u_{\alpha} u_{\beta}
\right ] 
\label{apmass}
\end{equation}
\begin{eqnarray}
\partial_t(n u_{\alpha}) & + & \partial_{\beta}(n u_{\alpha} u_{\beta})  = 
-\partial_{\alpha}p^w + \kappa n \partial_{\alpha}(\nabla^2 n)
\rule{0mm}{7mm}\label{apmom} \\
& + & \nu_{ap} 
\partial_{\beta}\left[n\left(\partial_{\alpha}u_{\beta}+
\partial_{\beta}u_{\alpha}\right)\right]
\rule{0mm}{6mm}\nonumber\\
& + &  \chi c^2 (\psi - \phi) 
\partial_{\beta}
\left[ \partial_{\alpha} (n u_{\beta}) + u_{\beta} \partial_{\alpha} n
+ u_{\alpha} \partial_{\beta} n
\right]
\rule{0mm}{6mm}\nonumber
\end{eqnarray}
Thus, the finite difference scheme introduces spurious terms, depending
on the quantity $(\psi - \phi)$, in both the conservation equations 
(compare with Eqs.~(\ref{physmass})-(\ref{physmom})),
while the physical value (\ref{physnu}) of the kinematic viscosity is replaced
by the {\emph{apparent value}} \cite{sosek2003}
\begin{equation}
\nu_{ap} = \chi c^2 (\tau+\psi)
\label{apnu}
\end{equation}
One could use $\delta s$, $\delta t$, and $c$ such that $\psi=\phi
\Leftrightarrow \delta s = c \delta t$ and
remove spurious terms in the Eqs.~(\ref{apmass})-(\ref{apmom}). 
In this case it would be 
$\nu_{ap} = \chi c^2 (\tau + \delta t / 2)$. 
In order to maintain the apparent value of the viscosity close to the physical
one and allow very small values of $\nu$, one should require 
$\delta t \ll \tau$. Since the condition $\psi=\phi$ is equivalent to ask
$N = L / c \delta t$, one should have $N \stackrel{>}{\sim} 10^{4}$ 
when $\tau \stackrel{<}{\sim} 10^{-3}$,
being $ c \simeq L \simeq 1$.
This would require a huge
computational effort when doing $2D$ or $3D$ simulations
using the first order upwind FDLB model. Higher order flux limiter schemes
provide a possibility to overcome this problem giving a better stability.
As we will see further, these schemes improve the accuracy of the FDLB
simulations with respect to the upwind scheme, for the same value of 
the number $N$ of lattice nodes.

As a matter of comparison we recall that the ``collide and stream'' LB model
is equivalent to an upwind FDLB model, when also the relaxation term is 
calculated on the characteristics line \cite{sosek2003} and
the choice $\delta s = c \delta t$ is adopted.
The resulting apparent value of the viscosity is 
$\nu_{ap} = \chi c^2 (\tau - \delta t / 2)$.
For this reason the ``collide and stream'' LB model 
suffers mainly from the lack of stability 
when $\tau \simeq \delta t / 2$ so that very low values of viscosity cannot
be accessed \cite{theofanous2003}.

\subsection{Flux limiter schemes}

\begin{figure}
\begin{tabular}{ccc}
\includegraphics[width=0.42\columnwidth]{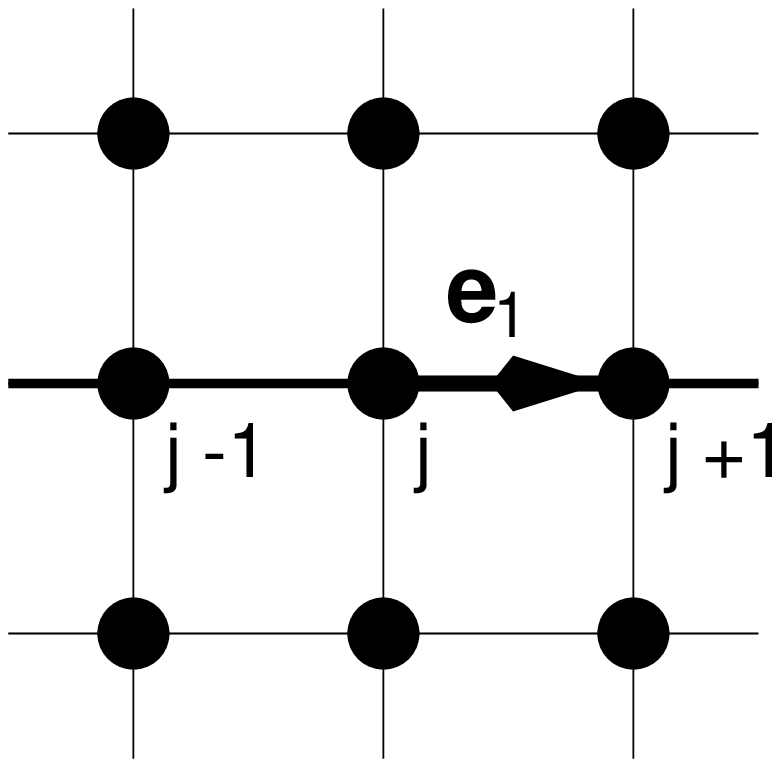}
& \qquad\qquad &
\includegraphics[width=0.42\columnwidth]{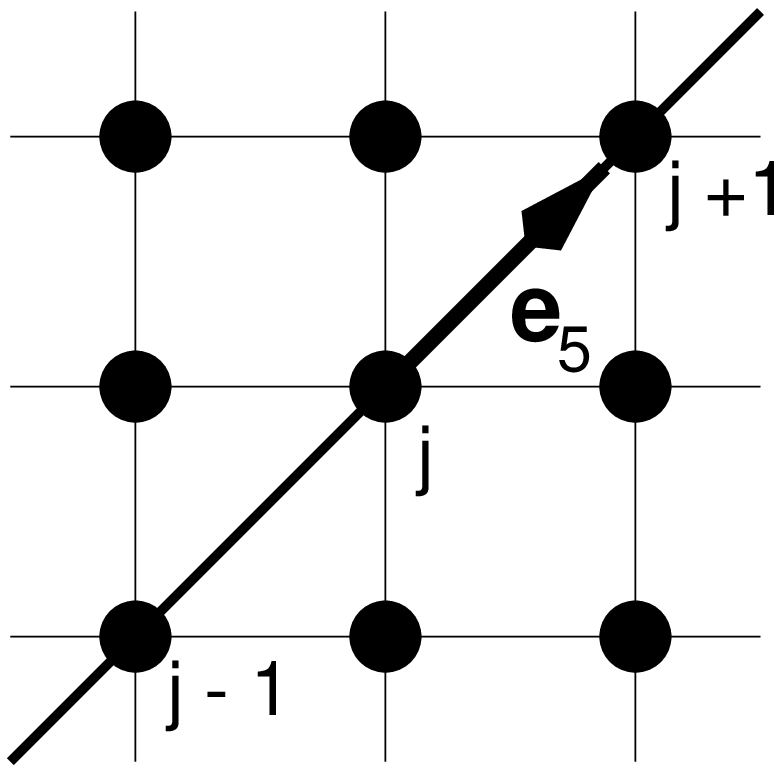}
\\ (a) & & (b)
\end{tabular}
\caption{Characteristics lines on the square lattice, for the
directions ${\bm{e}}_1$ (a) and ${\bm{e}}_5$ (b).\label{fchar}}
\end{figure}

Figure \ref{fchar} shows two characteristics lines on the square lattice
involving the distribution functions $f_{1}({\bm{x}},t)$ and
$f_{5}({\bm{x}},t)$, respectively. For convenience,
we denote $g_{i,j}^{k}$ the value
of the quantity $g_{i}$ in node $j$ at time
$t=k\delta t$. According to the general procedure to construct high
order Total Variation Diminishing (TVD) schemes using flux limiters
\cite{leveque,toro,leveque2002} we rewrite the updating rule (\ref{uprule})
in a conservative form using two fluxes \cite{teng,prorom,cejp}

\begin{equation}
f_{i,j}^{k+1} = f_{i,j}^{k} - \frac{c\delta t}{\delta s}
\left[ F_{i,j+1/2}^{k} - F_{i,j-1/2}^{k} \right] + \delta t Q_{i,j}^{k}
\label{conser}
\end{equation}
where
\begin{equation}
F_{i,j+1/2}^{k} = 
f_{i,j}^{k}\,+\,\frac{1}{2} \left[ 1 - \frac{c\delta t}{\delta s} \right]
\,\left[f_{i,j+1}^{k} - f_{i,j}^{k} \right]\,
\Psi(\Theta_{i,j}^{k})
\label{fluxdef}
\end{equation}
and
\begin{equation}
F_{i,j-1/2}^{n} = F_{i,(j-1)+1/2}^{n}
\end{equation}
The flux limiter $\Psi(\Theta_{i,j}^{n})$
introduced in (\ref{fluxdef}) is expressed as a function of the
{\emph{smoothness}}
\begin{equation}
\Theta_{i,j}^{n} =
\frac{f_{i,j}^{n} - f_{i,j-1}^{n}}
{f_{i,j+1}^{n} - f_{i,j}^{n}}
\label{smoothness}
\end{equation}
In particular, the second order Lax - Wendroff
scheme is recovered for $\Psi(\Theta_{i,j}^{n})\,=\,1$.
The upwind scheme, described in the previous subsection,
 is recovered as another
particular case, when $\Psi(\Theta_{i,j}^{n})\,=\,0$.
A wide choice of flux limiters are at our disposal in the
literature \cite{leveque,toro,leveque2002}. LB simulations reported
in this paper were done using the Monitorized Central Difference (MCD) limiter
\cite{leveque}
\begin{equation}
\Psi(\Theta_{i,j}^{n})\,=\,\left\{\begin{array}{ccl}
0 & , & \Theta_{i,j}^{n} \le 0 \\
2\Theta_{i,j}^{n} & , & 0 \le \Theta_{i,j}^{n} \le \frac{\,1\,}{\,3\,}
\rule{0mm}{10mm}\\
\frac{{\displaystyle{ 1 + \Theta_{i,j}^{n} }}}
{{\displaystyle{ 2 }}} & , &
\frac{1}{3} \le \Theta_{i,j}^{n} \le 3 \rule{0mm}{10mm} \\
2 & , & 3 \le \Theta_{i,j}^{n} \rule{0mm}{10mm}
\end{array}\right.
\label{mcd}
\end{equation}
but other limiters give qualitatively similar results.

Eqs.~(\ref{conser}) satisfy the global particle and momentum conservation.
When using the first order upwind scheme, the spurious terms 
introduced in the mass and momentum conservation
equations are linearly dependent on the lattice spacing $\delta s$. Since flux
limiter schemes are adapting
themselves to the local smoothness (\ref{smoothness}) of the distribution
functions, it is rather cumbersome to derive analytical expressions of
the spurious numerical term $\psi$ in these cases. LB simulations of
diffusion phenomena done using flux limiter schemes suggest a
second order dependence of the value $\psi$
on the lattice spacing $\delta s$ \cite{diffusivity}
such that $\psi = (\delta s)^2 / 2 c L$ and the apparent value of
the kinematic viscosity (\ref{apnu}) becomes
\begin{equation}
\nu_{ap\_flux}\,= \,\chi c^2 \left[\,\tau +
\frac{(\delta s)^2}{2cL}\,\right]
\label{apdif2}
\end{equation}
When the lattice spacing is a small quantity,
the use of flux limiter schemes is expected to improve the accuracy of
FDLB simulations as well as the stability. 

A different approach that allows to avoid spurious terms in the conservation 
equations is provided by the volumetric LB scheme introduced in 
Ref.~\cite{chen1998} 
which satisfies detailed balance and achieves the desired 
order of accuracy. We will refer to the fractional 
version of the aforementioned scheme constructed in the case of a
homogeneous fluid on a 
uniform mesh \cite{chen2001}
since we are using a regular and uniform lattice. 
In the scheme proposed in Ref.~\cite{chen2001} 
the value of the viscosity can 
be reduced with respect to the "collide and stream" LB and the 
Courant-Friedrichs-Levy number $CFL = c \delta t / \delta s$ can be smaller 
than 1. Moreover, some unphysical spurious invariants are removed. Our scheme 
can have very small values of viscosity since the numerical contribution to 
the value of viscosity, proportional to $\psi$, can be reduced and made much 
smaller than the physical term, proportional to $\tau$, without having 
stability problems. This depends on the fact the in 
finite difference schemes the values of $\delta s$ and $\delta t$ 
can be set independently from
the value of $c$. 
Our choice of $\delta t$ and $\delta s$ is such to guarantee that 
the $CFL$ number is much less than 1 and that the unavoidable spurious terms, 
introduced by the numerical scheme and proportional to $(\psi - \phi)$, can be 
done as small as desired.

\section{\label{secresults}Simulation results}

In this Section we report the results of our simulations.
For all runs we used $N=1024$, 
$\delta s = 1/256$, and $\delta t = 10^{-5}$.
In the following, lengths are expressed in units of lattice spacing 
and time is expressed as the product 
of algorithm steps by $\delta t$. 
All quenches below the critical temperature were to the 
temperature $\theta = 0.79$
where the coexisting densities are $n_{liquid} = 1.956$ and 
$n_{vapor} = 0.226$.
Each simulation was started with small fluctuations ($0.1 \%$) 
in the density about
a mean value ${\hat n}$ that was either symmetric (${\hat n} = 1.09$, 
liquid fraction $\beta=0.5$) or slightly off-symmetric (${\hat n} = 1.0$, 
$\beta=0.45$). The parameter $\kappa$ controlling the
surface tension was set to $5 \times 10^{-6}$ to have an interface 
thickness of $ \sim 6 $ lattice spacings. 
The viscosity was varied by changing $\tau$. We fixed an upper bound of
$\tau$ by the following argument.
It is well known that the continuum hypothesis and the
Navier - Stokes equation are valid only for small values of the Knudsen number
$Kn$ \cite{beskok}. Since $Kn = c\tau / L$ and $c \simeq L \simeq 1$ in our
simulations, this means $\tau \stackrel{<}{\sim} 10^{-2}$.
We implemented the upwind and the flux limiter schemes and compared results
when $\tau = 10^{-4}$. In this case the spurious numerical contribution of
the upwind scheme is 
larger than the physical one. Numerical contributions get negligible
when the  flux limiter scheme is considered instead. Therefore
one expects to observe qualitative and quantitative differences. 
We used also the value $\tau = 10^{-3}$ with the flux limiter scheme
to access a higher viscosity regime. 

In order to have different and independent tools to estimate the
domains size we used the following quantities:
$R_1(t)$, the inverse of the length of the interfaces of domains,
measured by counting lattice points where the 
order parameter $\rho({\bm x},t) = n({\bm x},t) - {\hat n}$ is such that 
$\rho({\bm x},t) \rho({\bm x'},t)< 0$;
$R_2(t)$, the inverse of the first moment of the spherically averaged
structure factor 
\begin{equation}
R_2(t) = \pi \frac{\int C(k,t) dk}{\int k C(k,t) dk} ,
\end{equation}
where $k = |{\bm k}|$ is the modulus of the wave vector in Fourier space, and
\begin{equation}
C(k,t) = < {\tilde \rho}({\bm k},t) {\tilde \rho}(-{\bm k},t) >
\end{equation}
with ${\tilde \rho}({\bm k},t)$ the spatial Fourier transform of the 
order parameter $\rho({\bm x},t)$.
The angle brackets denote an average over a shell in ${\bm k}$ space at fixed
$k$. The last quantity 
$R_3(t)$ is defined as
the  inverse of the first moment of the spherically averaged
structure factor of the fluid velocity 
\begin{equation}
R_3(t) = \pi \frac{\int C_u(k,t) dk}{\int k C_u(k,t) dk} 
\end{equation}
with $C_u(k,t)=< |{\bm {\tilde u}}({\bm k})|^2 >$.
In all the figures $R_1$ was multiplied by $4,000,000$ to be shown
in the same plot with $R_2$ and $R_3$.

\begin{figure}
\includegraphics[width=0.9\columnwidth]{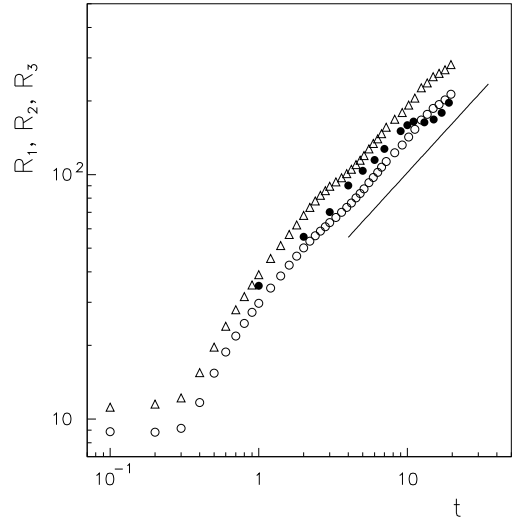}
\caption{Evolution of domains size recovered for
$\tau = 10^{-4}$, $\beta=0.5$ with the flux limiter scheme:
$R_1$ ($\triangle$), $R_2$ ($\circ$), $R_3$ ($\bullet$).
$R$'s are measured in lattice spacings and $R_1$ has been multiplied by
4,000,000 to be shown in the same plot. The straight line has slope $2/3$.
}
\label{figraggifllow}
\end{figure}

\begin{figure}
\begin{tabular}{ccc}
\includegraphics[width=0.43\columnwidth]{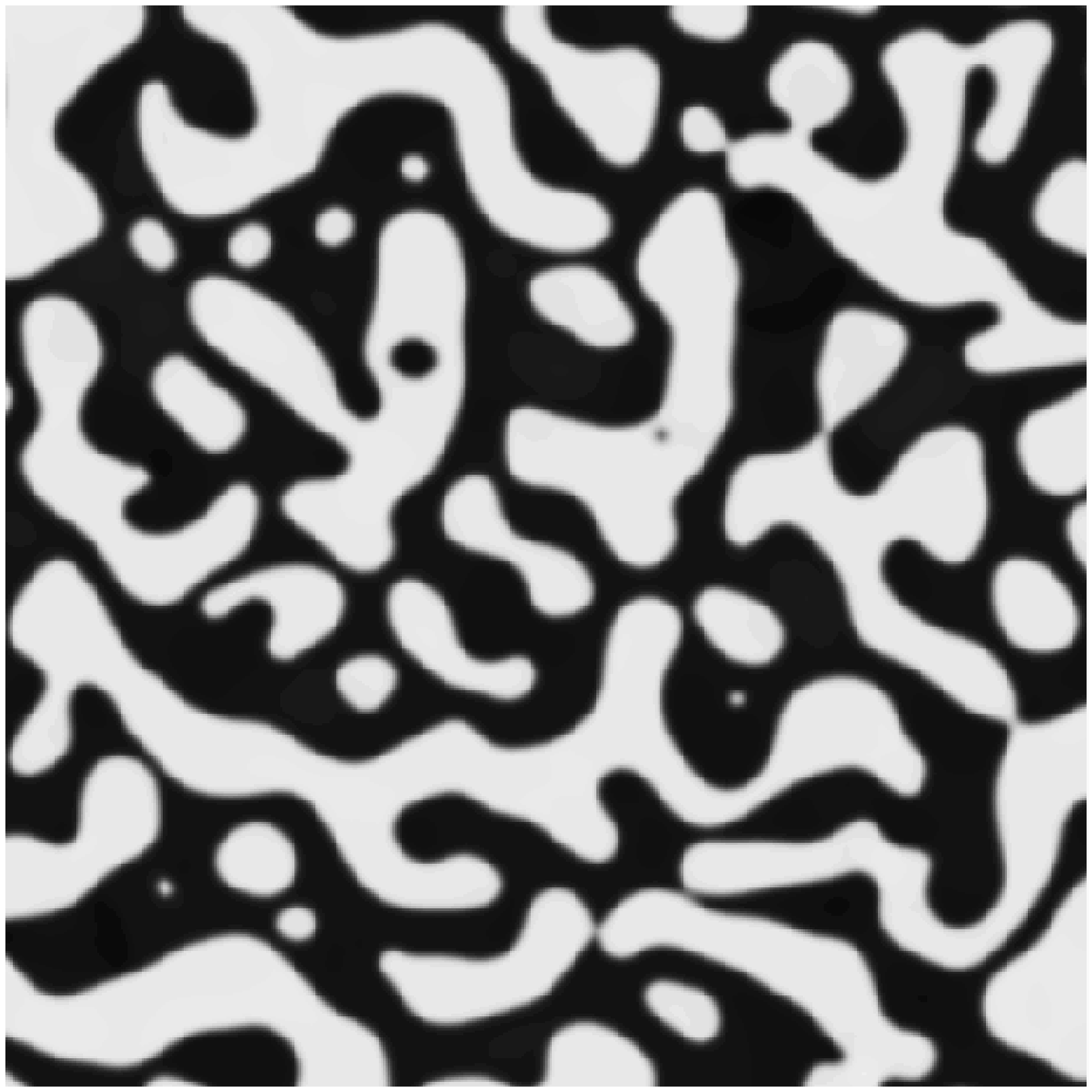}
& \qquad\qquad &
\includegraphics[width=0.43\columnwidth]{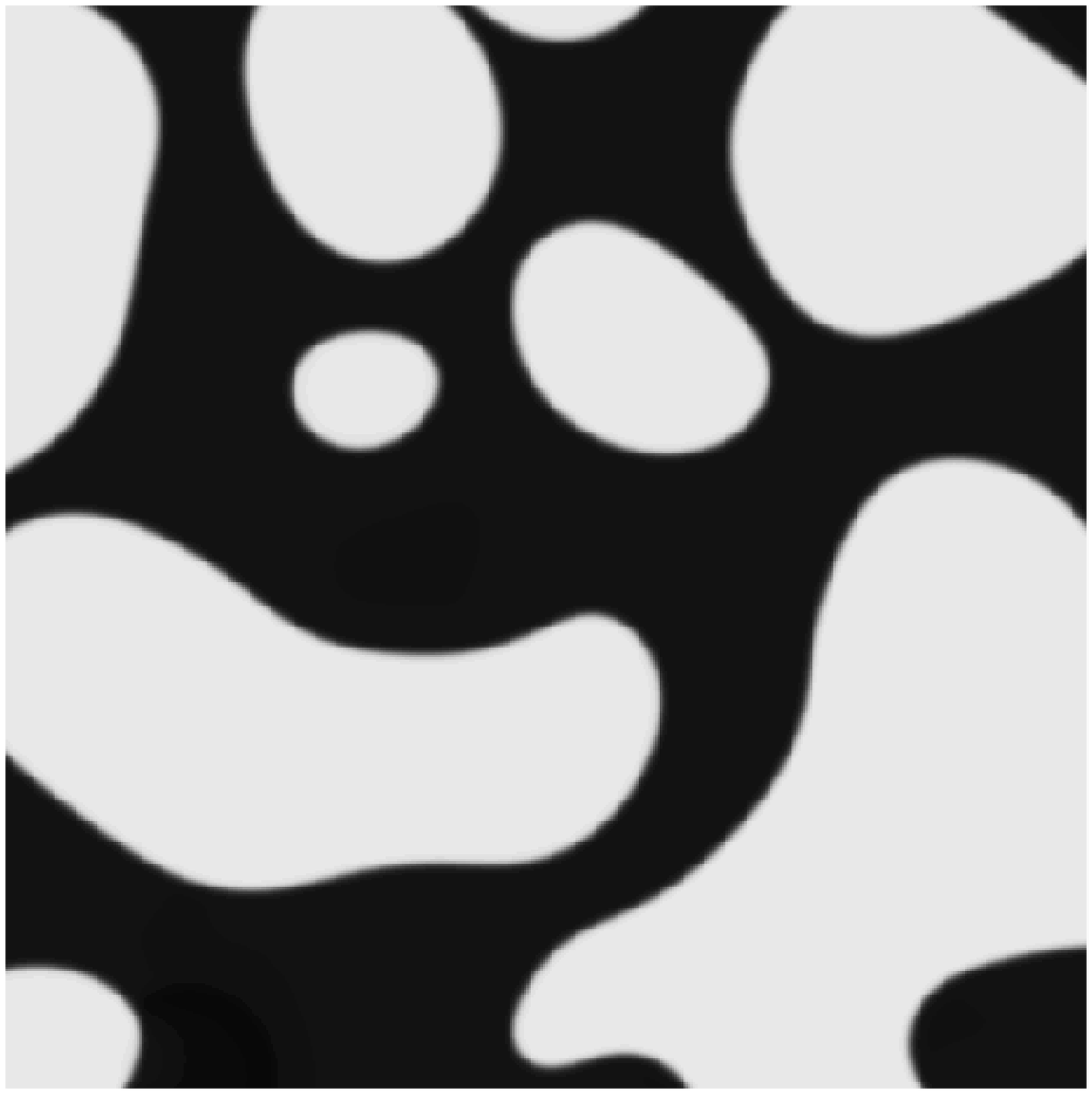}
\\ t=1 & & t=6\\
\includegraphics[width=0.43\columnwidth]{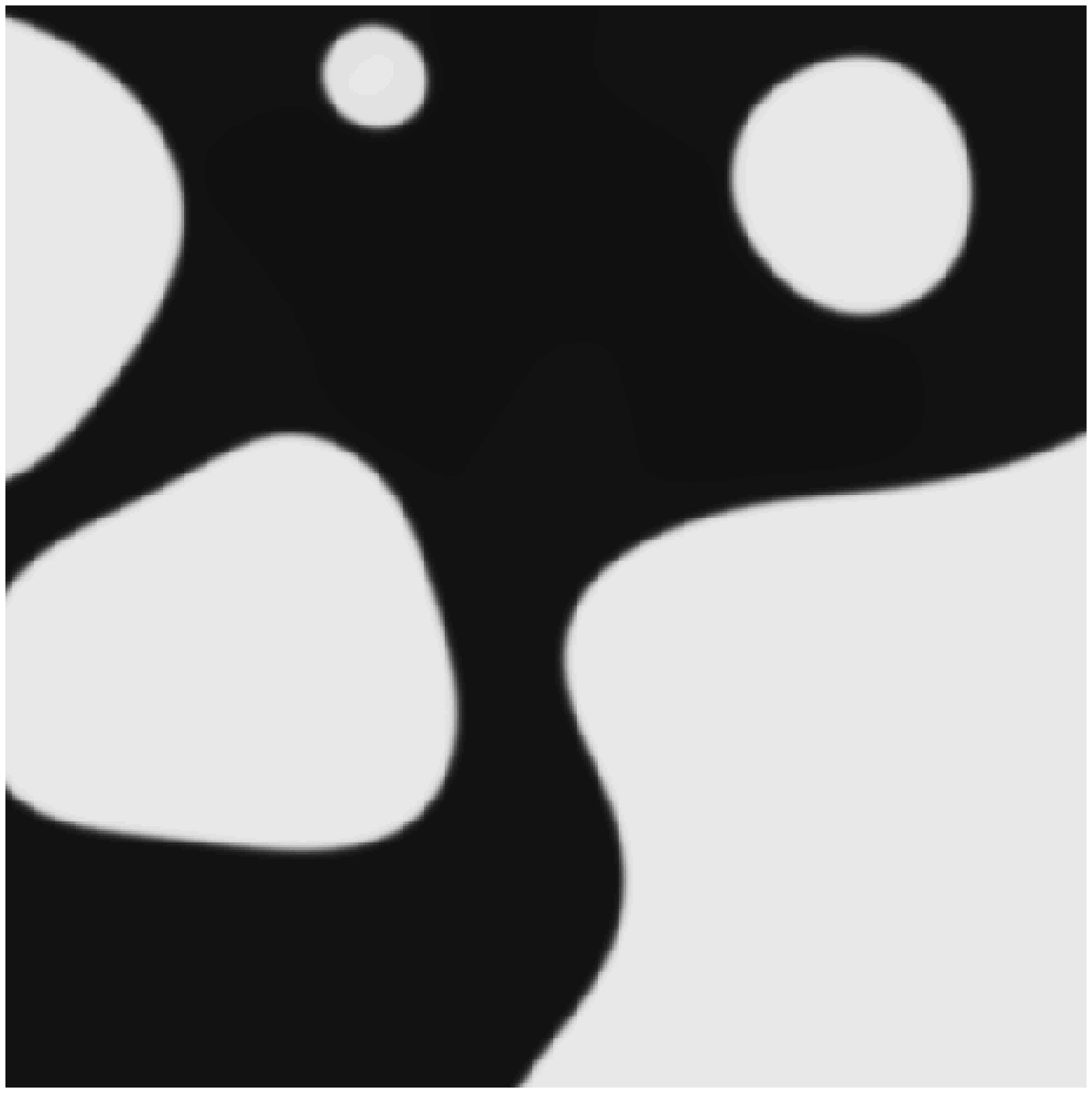}
& \qquad\qquad &
\includegraphics[width=0.43\columnwidth]{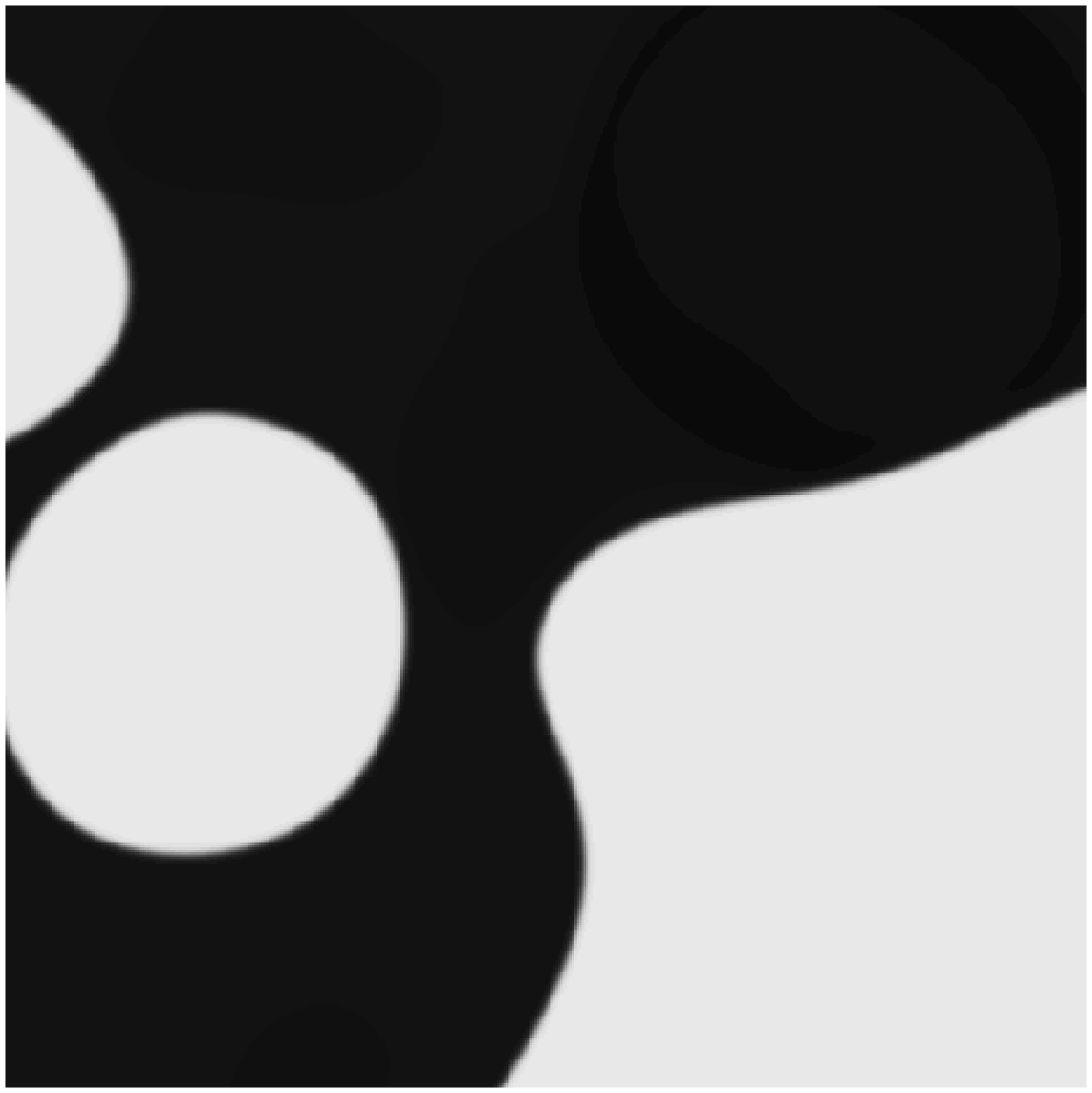}
\\ t=12 & & t=15
\end{tabular}
\caption{Contour plots of a portion $512 \times 512$ of the whole lattice
of the density $n$ in the case with $\tau = 10^{-4}$, $\beta=0.5$
 and flux limiter
scheme. Color code: black/white $\rightarrow$ liquid/vapor.}
\label{figplotfllow}
\end{figure}

In Fig.~\ref{figraggifllow} we present the three measures of domains size as
function of time for the 
case $\tau=10^{-4}$ with flux limiter 
scheme, and symmetric composition. It is 
interesting to note that $R_2$ and $R_3$ have the same trend, with a 
similar prefactor. This feature holds for all the runs we considered.
This last point is not obvious {\it a priori}.
After a swift initial growth
the evolution of
all quantities suggests the existence of the growth exponent $2/3$.
This is in accordance with previous studies on symmetric 
liquid-vapor systems at low viscosity \cite{osborn} when hydrodynamic flow 
is operating. In this regime hydrodynamics is the mechanism to get domains
circular since the flow is driven by the difference in Laplace pressure
between points of different curvature on the boundary of domains. 
This remark is confirmed when looking at configurations of the density $n$.
In Fig.~\ref{figplotfllow} we show contour plots of a part of the whole
system at consecutive times. The vapor bubble in the down left corner at
$t=12$, while evaporating, 
is rounded by the flow as it can be seen by comparing it with the shape
at $t=15$.

\begin{figure}
\includegraphics[width=0.9\columnwidth]{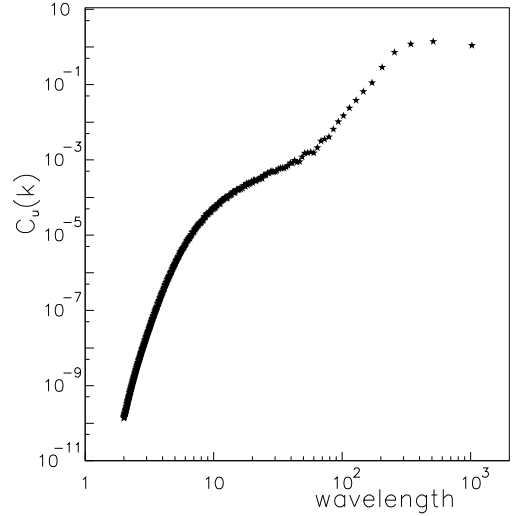}
\caption{Velocity structure factor $C_u(k)$ at time $t=15$ 
in the case with $\tau = 10^{-4}$, $\beta=0.5$  and flux limiter scheme.
$C_u(k)$ is in arbitrary units and the wavelength is measured in 
lattice spacings.}
\label{figfattstrfllow}
\end{figure}

An indication about the velocity field comes from the 
structure factor $C_u(k,t)$.
In Fig.~\ref{figfattstrfllow} we plot it at time $t=15$.  
It exhibits a structure at a scale comparable with system size.
All velocity components decay 
becoming small at low wavelengths and contributing 
little to the overall dynamics. 
A small bump can be seen at wavelength $ \sim 8$  corresponding to  
capillary motion at interface length scale.
A similar behavior was observed in binary fluids \cite{kendon}.

\begin{figure}
\includegraphics[width=0.9\columnwidth]{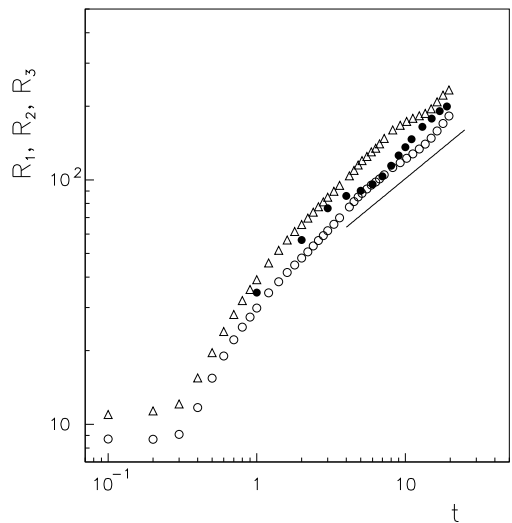}
\caption{Evolution of domains size
in the case with $\tau = 10^{-4}$, $\beta=0.5$ and the upwind scheme: 
$R_1$ ($\triangle$), $R_2$ ($\circ$), $R_3$ ($\bullet$).
$R$'s are measured in lattice spacings and $R_1$ as been rescaled by
4,000,000 to be shown in the same plot. The straight line has slope $1/2$.}
\label{figraggiuwlow}
\end{figure}

\begin{figure}
\includegraphics[width=0.95\columnwidth]{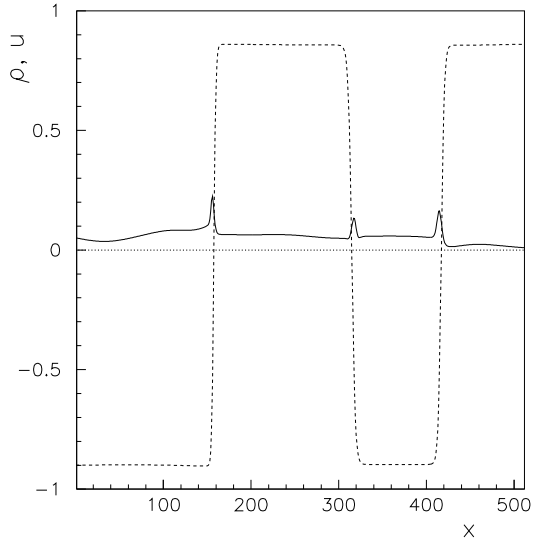}\\
\includegraphics[width=0.95\columnwidth]{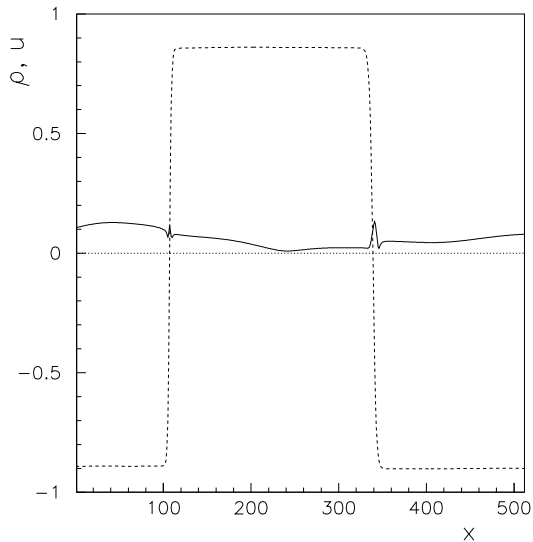}
\caption{Order parameter $\rho$ (dashed line) and velocity modulus $u$
(full line) profiles are shown along the line taken at $y=256$ lattice
spacings from bottom at time $t=20$ for the upwind scheme (upper panel) and
flux limiter scheme (lower panel) with $\tau=10^{-4}$, $\beta=0.5$.}
\label{figvellow}
\end{figure}

In the case with the upwind scheme the estimation of the growth
exponent is more difficult since data are noisy and none of the $R$'s
shows a clear trend.
From Fig.~\ref{figraggiuwlow} it seems the system enters a late
regime characterized 
by an exponent consistent with the value $1/2$.
We believe that this behavior is due to spurious 
terms in the macroscopic equations that are
considerably larger when using the upwind scheme than in the case with
flux limiter. These terms produce a numerical diffusivity when 
they are not negligible.
This is confirmed by the analysis of the velocity fields in the
two cases. In Fig.~\ref{figvellow} we plot the order parameter $\rho$
and velocity modulus $u$ along a horizontal 
cross section of the system taken at the same long time.
Two comments are in order here. It is quite unavoidable to have 
spurious velocities at interfaces where density gradients are present
with LB models (irrespectively of the particular model used 
\cite{houshan97}).
And also the present model shows this unpleasant feature. 
Nonetheless
it is evident that the flux limiter scheme allows to dump considerably
these spurious contributions. Indeed, with flux limiter the maximum 
value of velocity at interface is $ 0.13$ ($Ma = u / c_s = 0.14$)
while with the upwind
it is about 2 times larger being $ 0.23$ ($Ma = 0.26$).
The high value of the Mach number $Ma$
makes the expansions (\ref{feqdef}) 
less reliable with the upwind scheme.

\begin{figure}
\includegraphics[width=0.9\columnwidth]{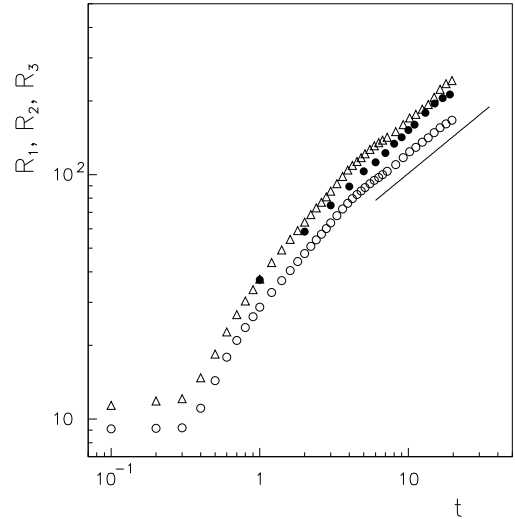}
\caption{Evolution of domains size
in the case with $\tau = 10^{-3}$, $\beta=0.5$ and flux limiter scheme: 
$R_1$ ($\triangle$), $R_2$ ($\circ$), $R_3$ ($\bullet$).
$R$'s are measured in lattice spacings and $R_1$ as been multiplied by
4,000,000 to be shown in the same plot. The straight line has slope $1/2$.}
\label{figraggiflhigh}
\end{figure}

\begin{figure}
\begin{tabular}{ccc}
\includegraphics[width=0.43\columnwidth]{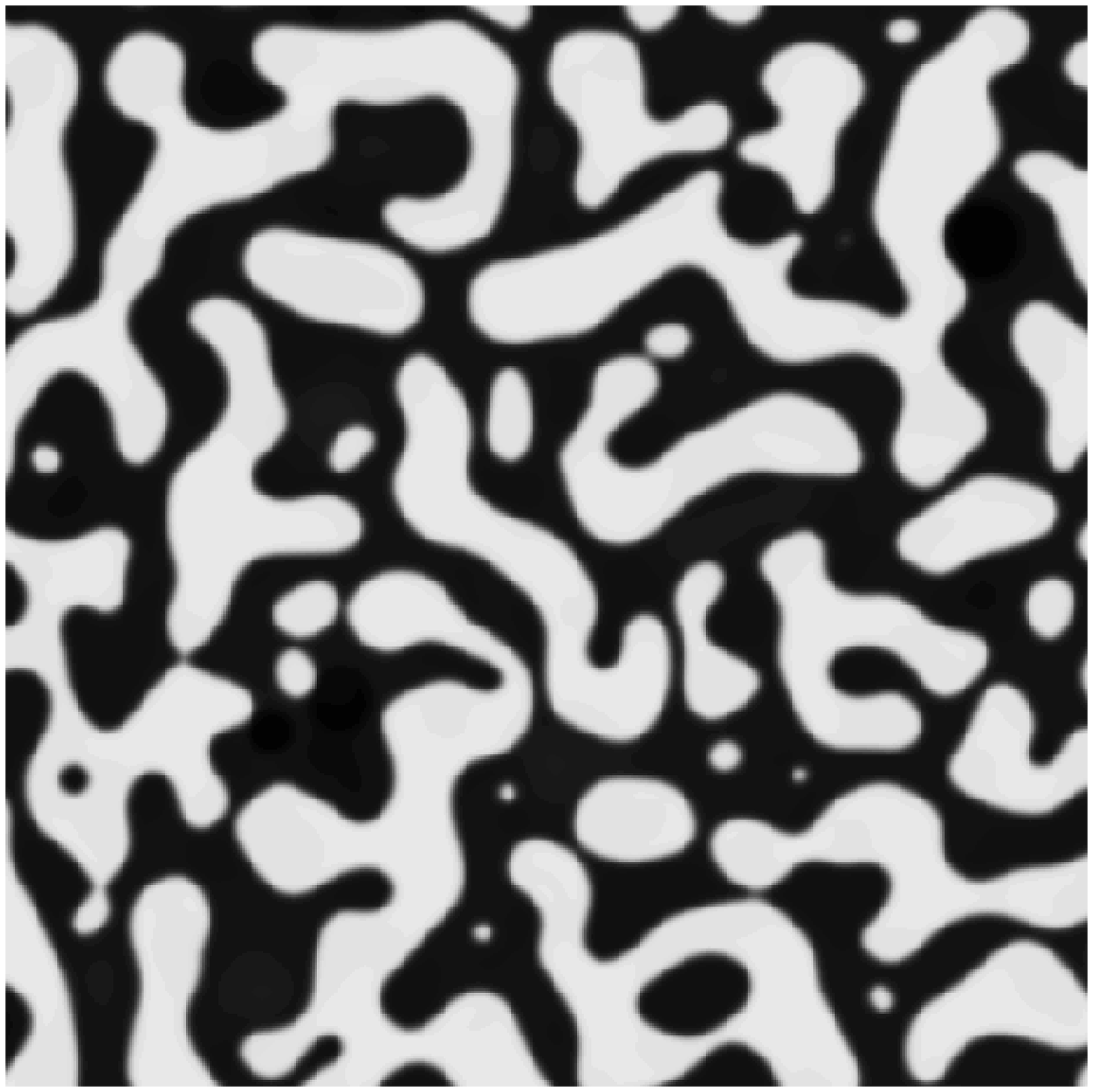}
& \qquad\qquad &
\includegraphics[width=0.43\columnwidth]{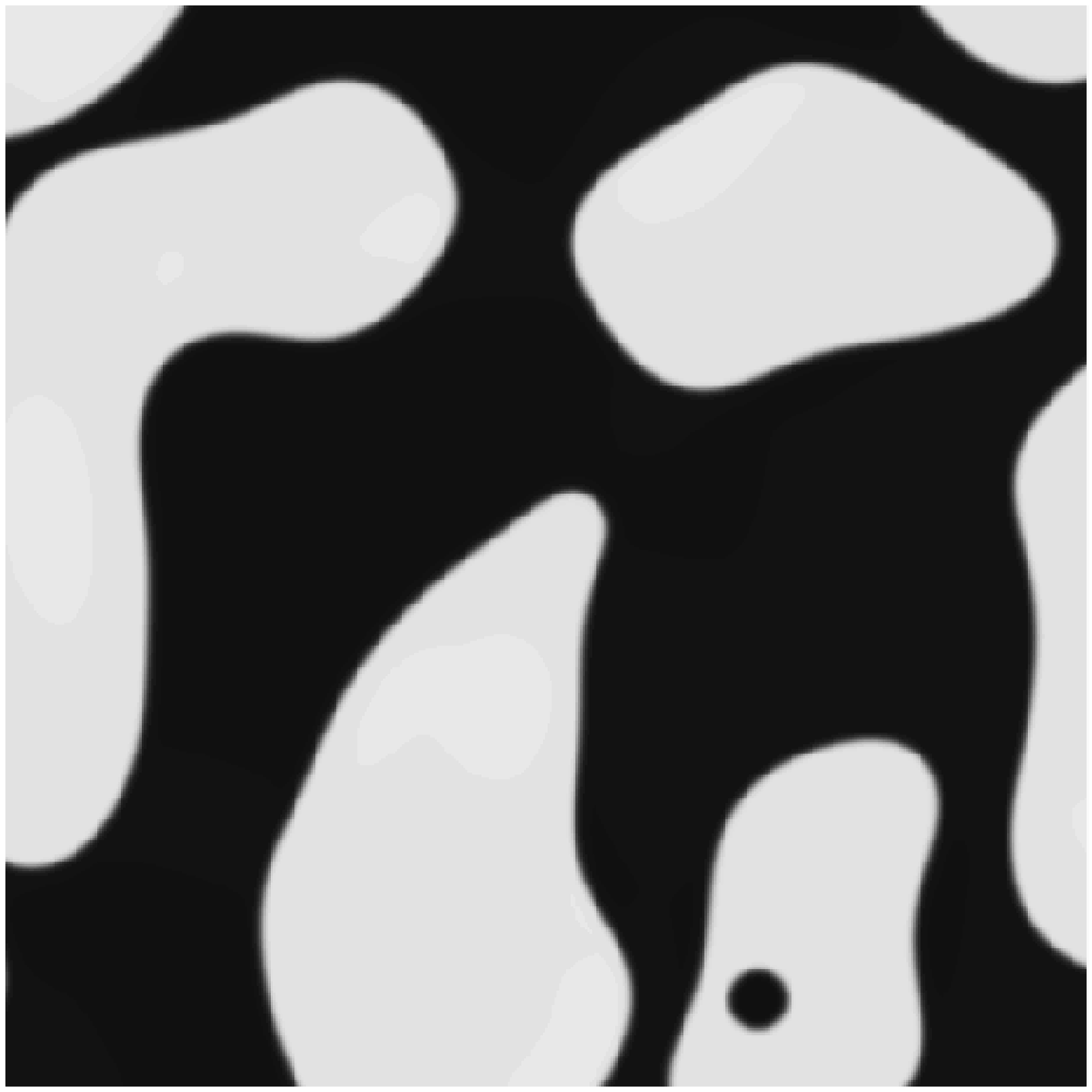}
\\ t=1 & & t=6\\
\includegraphics[width=0.43\columnwidth]{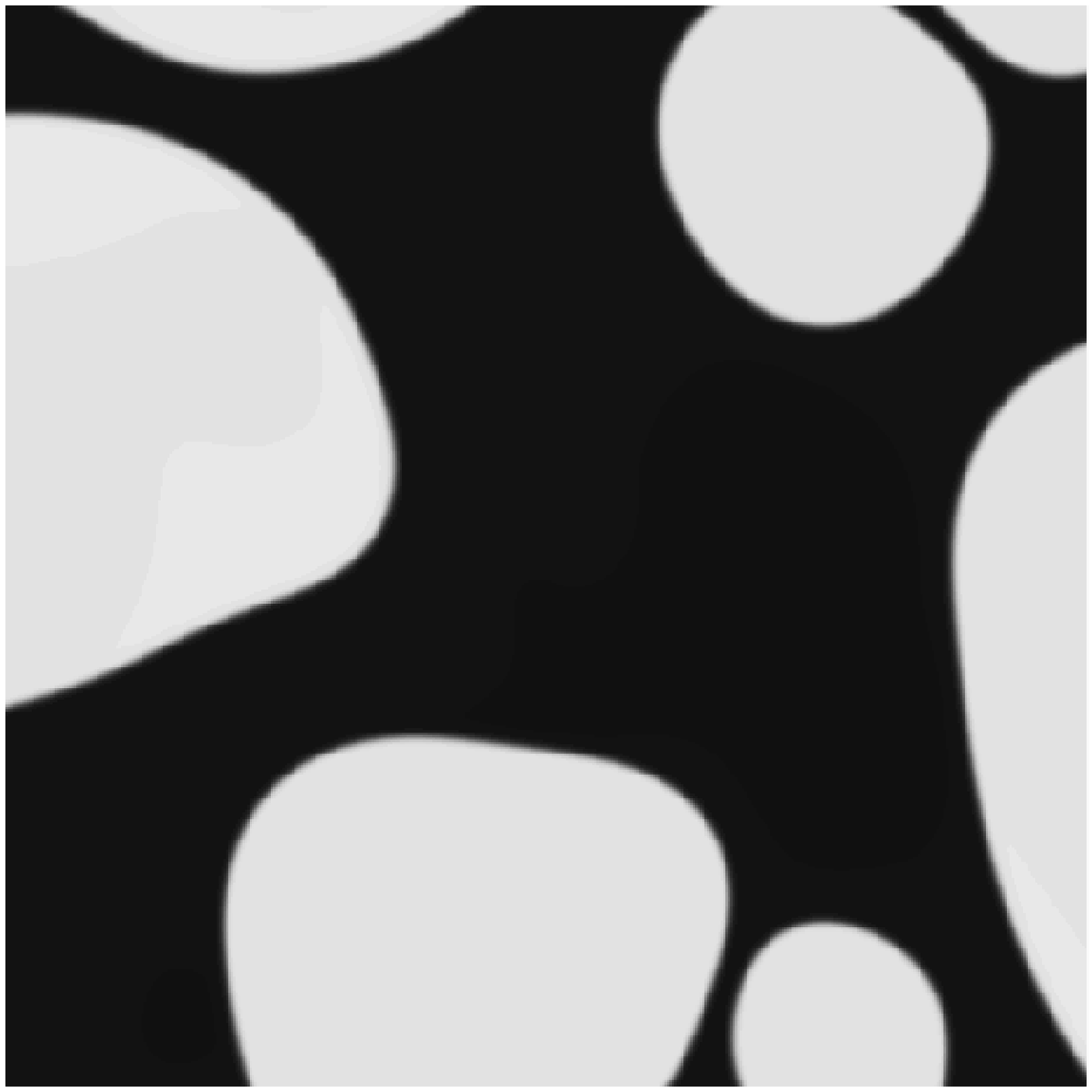}
& \qquad\qquad &
\includegraphics[width=0.43\columnwidth]{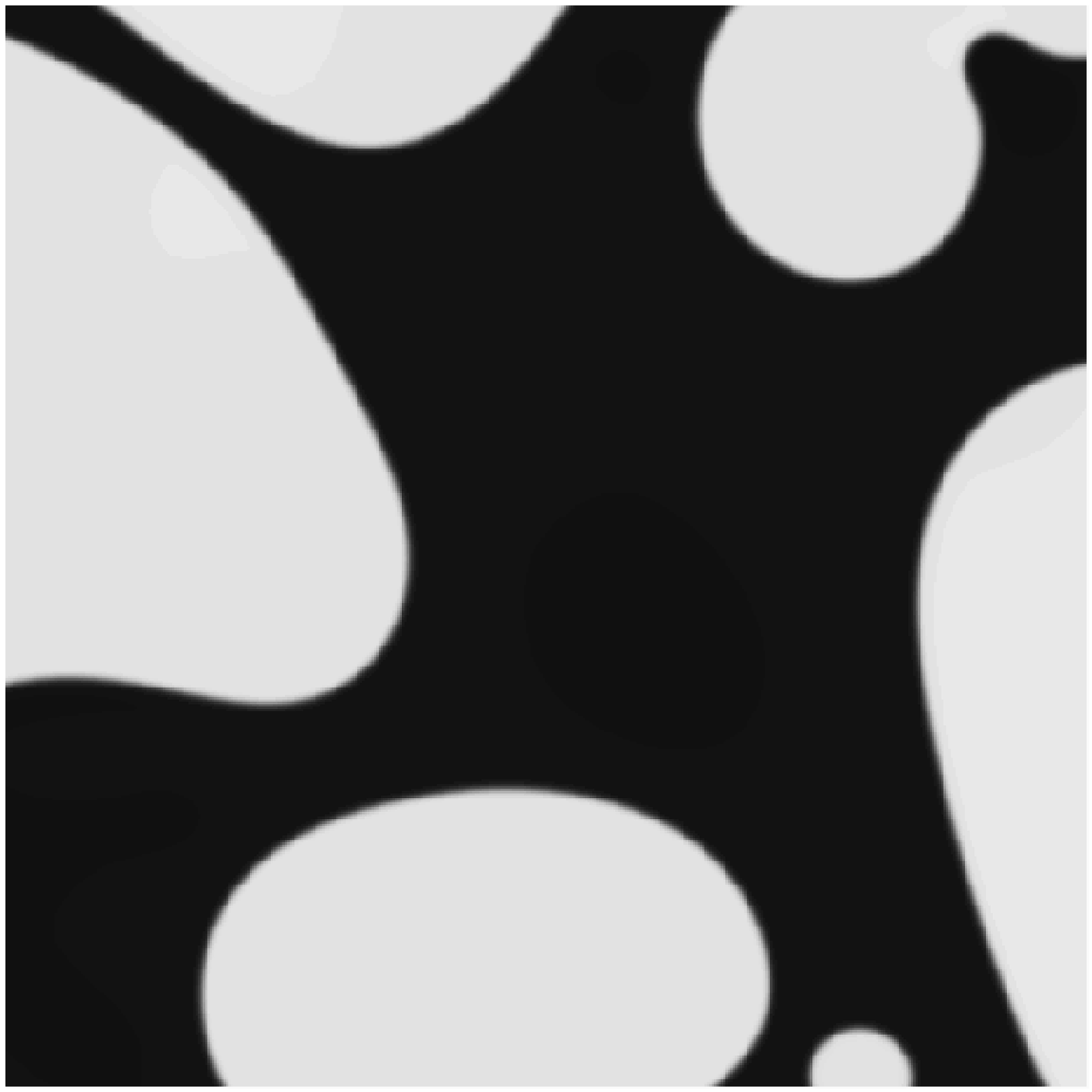}
\\ t=12 & & t=15
\end{tabular}
\caption{Contour plots of a portion $512 \times 512$ of the whole lattice
of the density $n$ in the case with $\tau = 10^{-3}$, $\beta=0.5$
 and flux limiter
scheme. Color code: black/white $\rightarrow$ liquid/vapor.}
\label{figplotflhigh}
\end{figure}

Due to the better performance of the flux limiter scheme we decided to adopt 
it for further simulations.
In Fig.~\ref{figraggiflhigh} we plot the three measures of domains size as
function of time for the 
case $\tau=10^{-3}$ with symmetric composition. 
After  initial growth
all the quantities suggest the existence of the growth exponent $1/2$.
This is in accordance with previous studies on liquid-vapor systems 
at high viscosity \cite{osborn} at symmetric composition 
when growth is expected to be described by the Allen-Cahn theory 
of interfaces dynamics which gives an exponent $1/2$ \cite{allen} 
and hydrodynamics is not operating.
Due to limits imposed by system size 
we cannot access very long times to probe whether
the hydrodynamic regime is the late regime as previously argued \cite{osborn}.
Fig.~\ref{figplotflhigh} shows density contour plots at consecutive times. 
Growth seems to be mainly driven by evaporation of vapor domains.

\begin{figure}
\includegraphics[width=0.9\columnwidth]{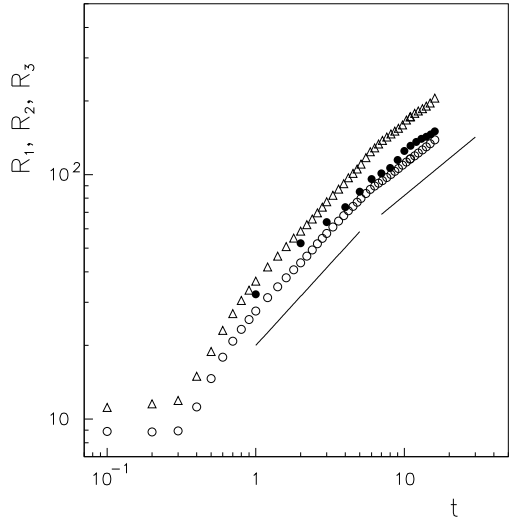}
\caption{Evolution of domains size
in the case with $\tau = 10^{-4}$, $\beta=0.45$ and flux limiter scheme: 
$R_1$ ($\triangle$), $R_2$ ($\circ$), $R_3$ ($\bullet$).
$R$'s are measured in lattice spacings and $R_1$ as been multiplied by
4,000,000 to be shown in the same plot. The straight lines have slopes
$2/3$ and $1/2$.}
\label{figraggifloff}
\end{figure}

\begin{figure}
\begin{tabular}{ccc}
\includegraphics[width=0.43\columnwidth]{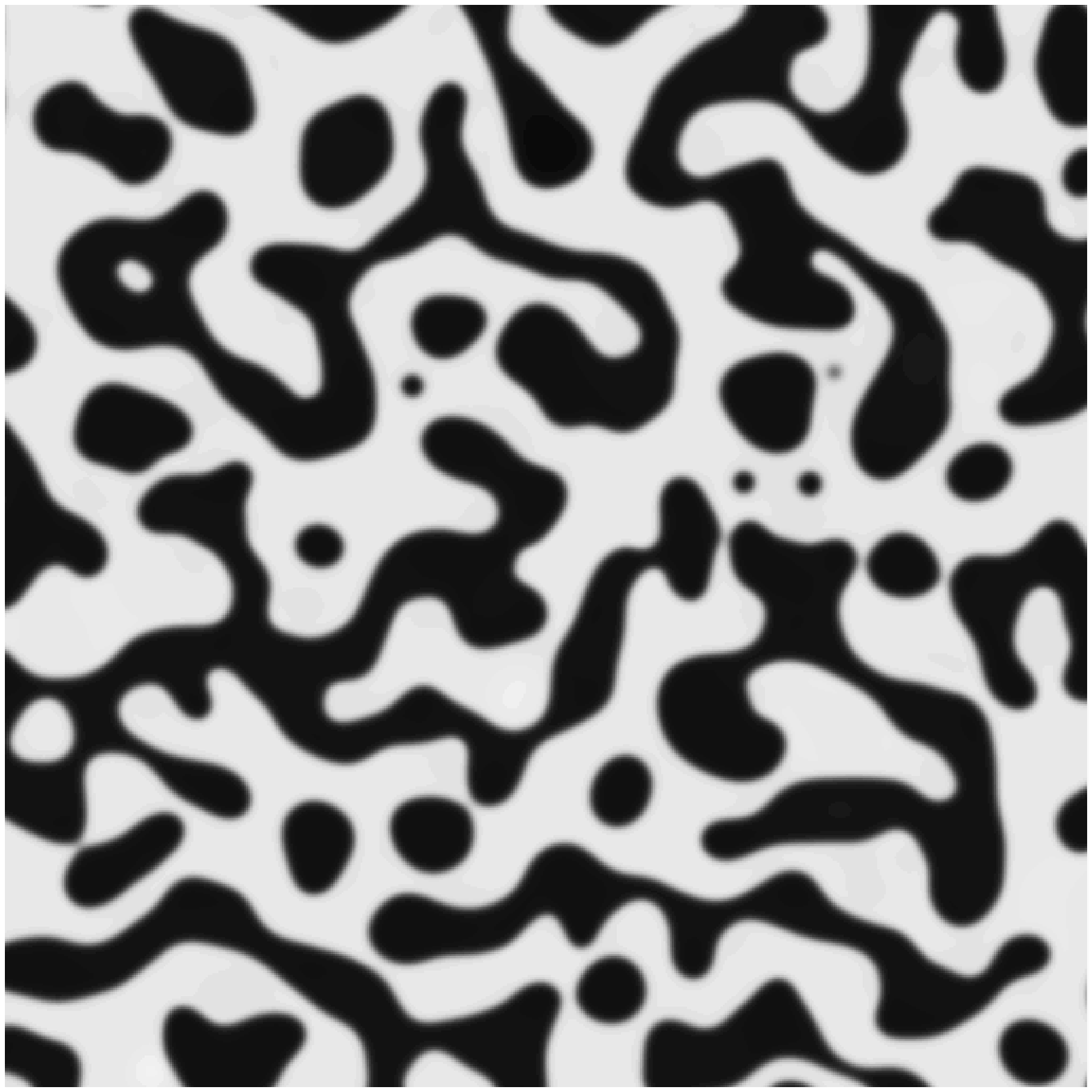}
& \qquad\qquad &
\includegraphics[width=0.43\columnwidth]{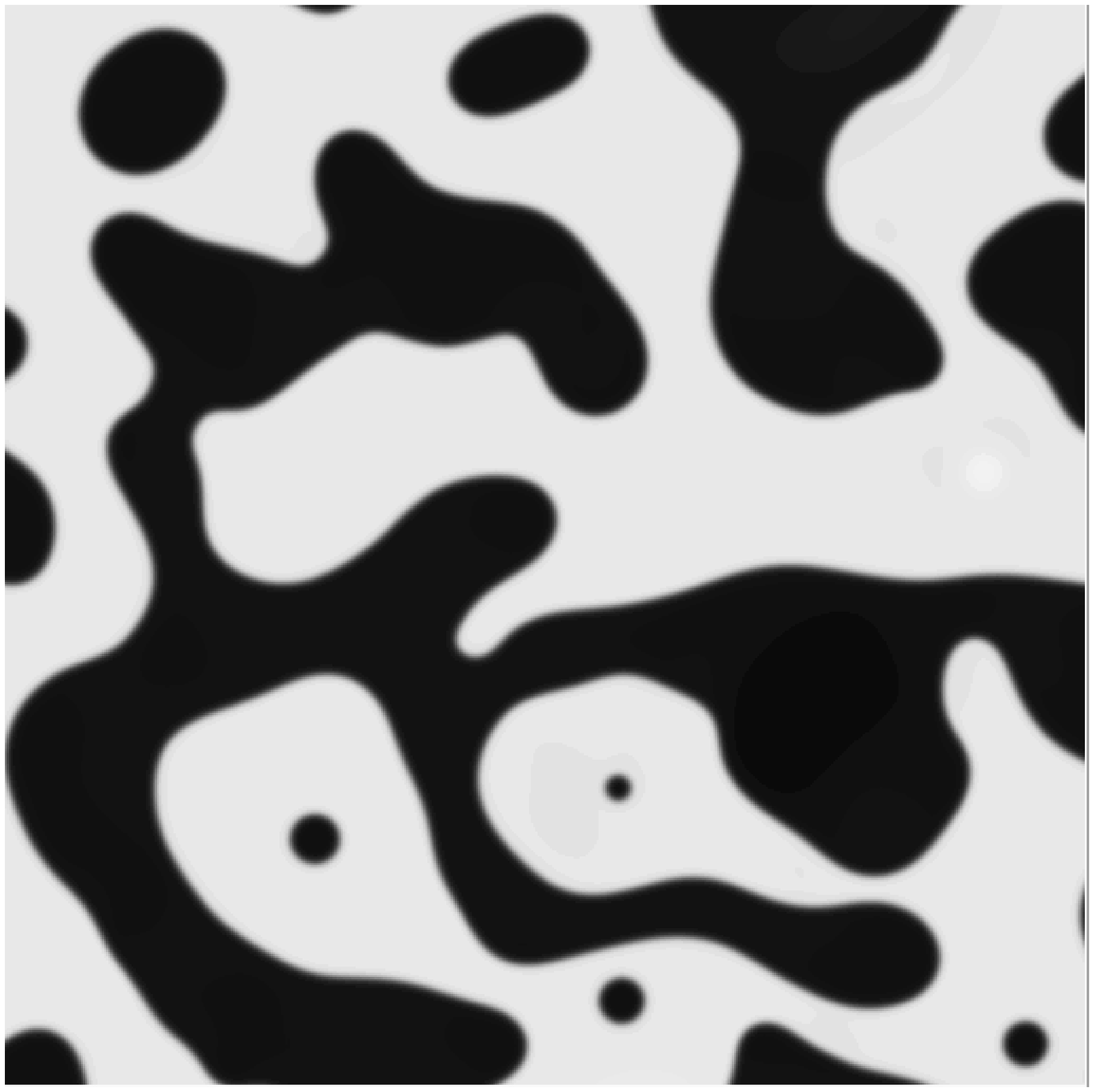}
\\ t=1 & & t=3\\
\includegraphics[width=0.43\columnwidth]{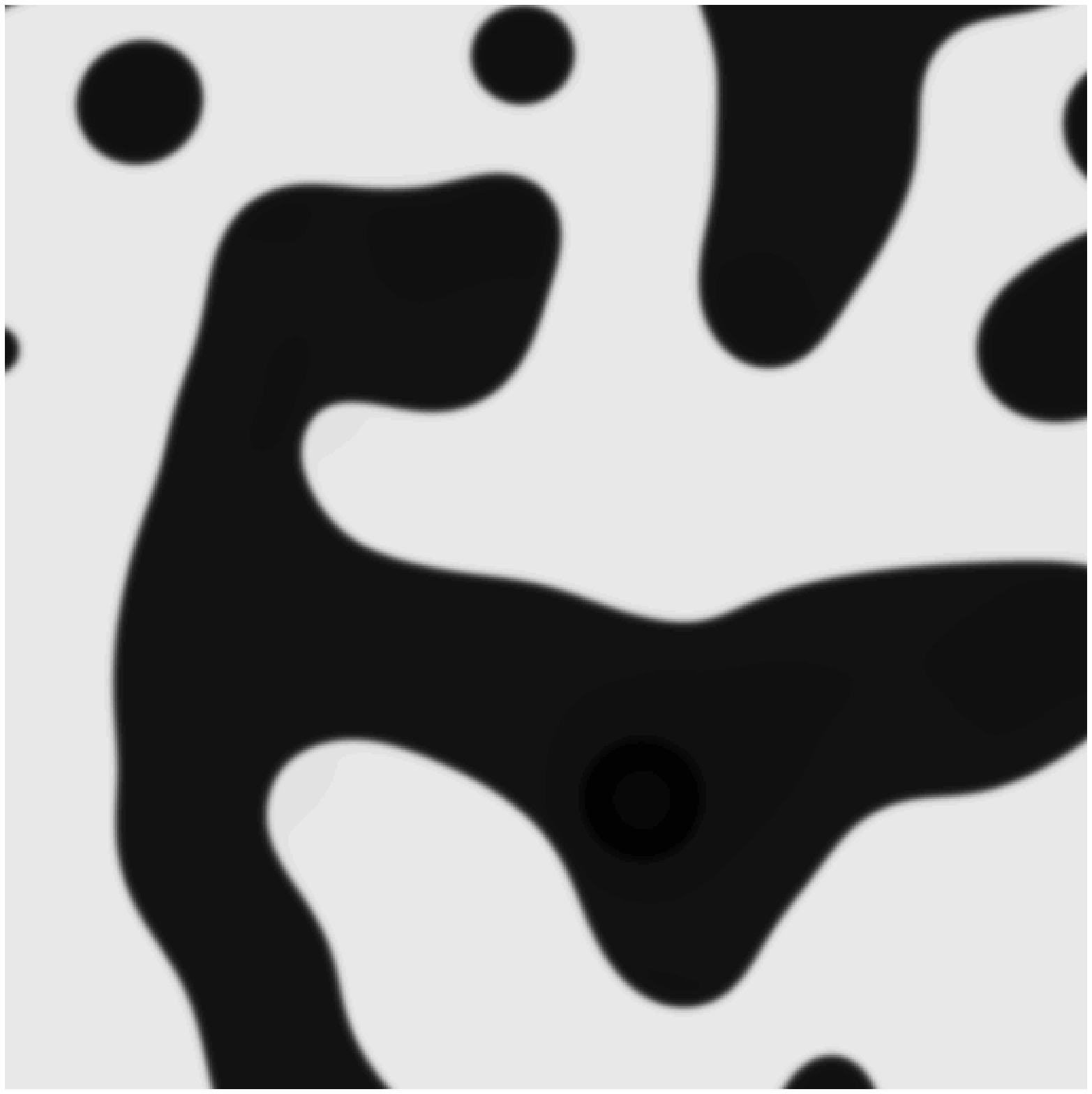}
& \qquad\qquad &
\includegraphics[width=0.43\columnwidth]{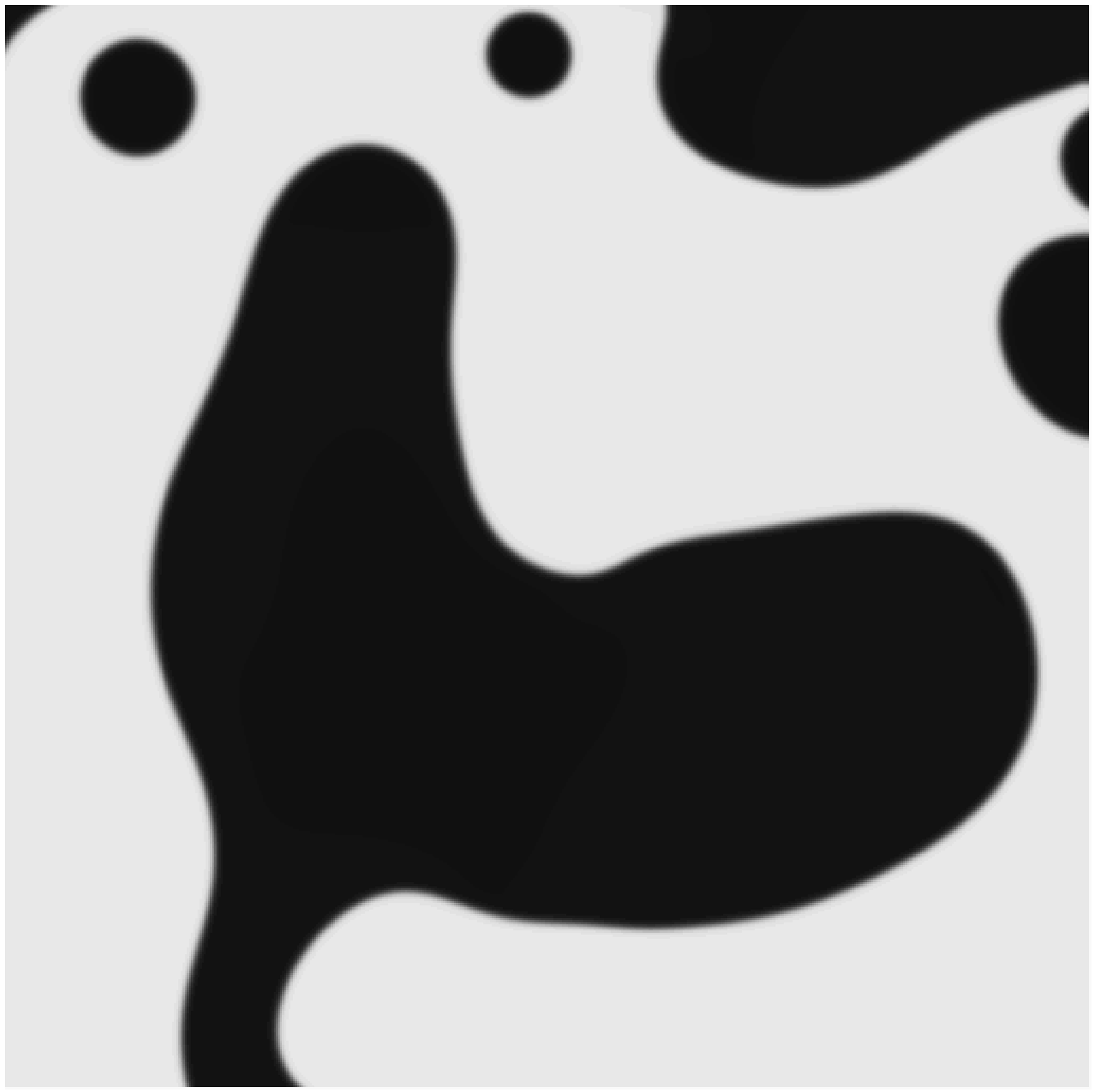}
\\ t=6 & & t=10
\end{tabular}
\caption{Contour plots of a portion $512 \times 512$ of the whole lattice
of the density $n$ in the case with $\tau = 10^{-4}$, $\beta=0.45$. 
Color code: black/white $\rightarrow$ liquid/vapor.}
\label{figplotfloff}
\end{figure}

Finally, we considered the case of an off-symmetric system with a liquid
fraction $\beta=0.45$.
In Fig.~\ref{figraggifloff} we plot the three measures of domains size as
function of time for the case $\tau=10^{-4}$.
After the initial growth
all the quantities suggest the existence of a
growth exponent $2/3$ which quite soon changes to $1/2$. 
In previous studies with a free-energy
LB model it was found the growth exponent to 
be $2/3$ with liquid fraction $\beta=0.31$ \cite{osborn} 
and $2/3$ crossing over to 
$1/2$ at $\beta=0.17$ \cite{mecke1}.
The problem of off-symmetric liquid-vapor
phase separation was recently addressed in Ref.~\cite{warren}. There
it was pointed out that in the case
of a dispersion of liquid drops in vapor, the growth should 
proceed with an exponent $1/2$ and the result 
was proven in the case of highly asymmetric composition with $\beta=0.1$. 
We believe that we are probing a regime similar to that seen
in two-dimensional binary fluids where, once hydrodynamics flow has made
domains circular, Allen-Cahn growth takes over \cite{yeom}. 
This interpretation seems to be confirmed by configurations of the
system, presented in Fig.~\ref{figplotfloff}. They
show that liquid drops in the vapor matrix are almost circular at $t=6$
so that the hydrodynamic mechanism is no more effective.

\section{Conclusions}

The correct choice of the numerical scheme is essential to recover 
the real physics of a fluid system subjected to LB simulations.
In the case of a liquid-vapor system we have seen that simulation results
exhibit significant changes when the numerical contribution of the
finite difference scheme to the apparent value of the transport
coefficients becomes comparable with the expected physical value.
The numerical contribution of the first order upwind scheme
is linearly dependent on the lattice spacing $\delta s$ and switches
to an higher order for the flux limiter scheme.
Since $\delta s < 1$, the flux limiter scheme
reduces the computing effort in terms of required lattice nodes
and gives physical results which are more accurate
for the same number of lattice nodes per unit length.  
Spurious velocities at interfaces can be considerably damped
and very low viscosity systems can be simulated preserving numerical stability.
The main limitation comes from the requirement of a small time step
when very low values of viscosity are needed.
To give an idea of the CPU time we report
that our code takes 6 hours to perform $10^5$ algorithm steps by using 
32 Xeon 3.055 GHz processors on the IBM Linux Cluster 1350 at CINECA 
\cite{cineca} with Myrinet IPC network 
and the Portable Extensible Toolkit for
Scientific Computation (PETSc 2.1.6) developed at Argonne National
Laboratory, Argonne, Illinois \cite{petsc}.

The model allowed to clarify the picture of phase separation in liquid-vapor
system. We found that the growth exponent depends on either
the fluid viscosity and the system composition. When liquid and vapor are 
present in the same amount, the growth exponent is $2/3$ and $1/2$ at low and
high viscosity, respectively. When the liquid fraction is less abundant than 
the vapor one, we can access a late time regime at low viscosity.
In this regime
the hydrodynamic transport is no more effective so we are able to see
the crossover from the exponent $2/3$ to $1/2$ which is characteristic
of the Allen-Cahn growth mechanism.

Finally, we note that our results as well as previous ones have been 
obtained in the case of isothermal systems. It would be interesting
to incorporate the energy conservation equation into the model 
to allow non uniform temperatures in the 
liquid-vapor system undergoing phase separation.

\begin{acknowledgments}
V.S. acknowledges financial support from the University of Bari and thanks
the IAC - CNR, Sezione di Bari, for hospitality.
Computer runs were done on the parallel computing clusters at
the Department of Physics, University of Bari and at the CINECA Consortium
for Supercomputing \cite{cineca} in Casalecchio di Reno (Bologna) under an
INFM grant. This work has been partially supported by MIUR (PRIN-2002).
\end{acknowledgments}

\bibliography{lamuraetal_2}

\end{document}